\documentclass[12pt]{article}
\usepackage{graphicx}
\usepackage{dcolumn}
\usepackage{bm}
\usepackage{amsmath, amssymb, amsfonts, amsthm}
\usepackage{cite}
\usepackage{epsfig}
\usepackage{textcomp}
\usepackage{tikz-cd} 
\usepackage{ dsfont }
\usepackage{amsmath}
\usepackage{color}

\def\a{\alpha}
\def\b{\beta}
\def\g{\gamma}

\def\f{\phi}

\def\m{\mu}
\def\n{\nu}
\def\r{\rho}
\def\s{\sigma}

\def\o{\omega}

\def\be{\begin{equation}}
\def\ee{\end{equation}}
\def\ba{\begin{eqnarray}}
\def\ea{\end{eqnarray}}

\def\bdm{\begin{displaymath}}
\def\edm{\end{displaymath}}

\def\bq{\begin{quote}}
	\def\eq{\end{quote}}

 at 10truept

\newcommand{\bea}{\begin{eqnarray}}
\newcommand{\eea}{\end{eqnarray}}

\newcommand{\bi}{\begin{itemize}}
	\newcommand{\ei}{\end{itemize}}

\newcommand{\Poincare}{Poincar\'{e}~}

\newcommand{\beq}{\begin{equation}}
\newcommand{\eeq}{\end{equation}}
\newcommand{\beqa}{\begin{eqnarray}}
\newcommand{\eeqa}{\end{eqnarray}}

\def\ltap{\ \raise.3ex\hbox{$<$\kern-.75em\lower1ex\hbox{$\sim$}}\ }
\def\gtap{\ \raise.3ex\hbox{$>$\kern-.75em\lower1ex\hbox{$\sim$}}\ }
\def\gl{\ \raise.5ex\hbox{$>$}\kern-.8em\lower.5ex\hbox{$<$}\ }
\def\roughly#1{\raise.3ex\hbox{$#1$\kern-.75em\lower1ex\hbox{$\sim$}}}

\begin{document}

		\thispagestyle{empty}
		\begin{flushright}
			January 2018\\
		\end{flushright}
		\vspace*{.2cm}
		\begin{center}
	
{\Large \bf Symmetry Breaking Patterns for Inflation}

			\vspace*{.7cm} {\large  Remko Klein\footnote{\tt
					remko.klein@rug.nl}, Diederik Roest\footnote{\tt d.roest@rug.nl
				} and  David Stefanyszyn\footnote{\tt
				d.stefanyszyn@rug.nl}}
		
		\vspace{.3cm} {\em Van Swinderen Institute for Particle Physics and Gravity, University of Groningen, Nijenborgh 4, 9747 AG Groningen, The Netherlands}\\

		\vspace{1cm} ABSTRACT 
	\end{center}
We study inflationary models where the kinetic sector of the theory has a non-linearly realised symmetry which is broken by the inflationary potential. We distinguish between kinetic symmetries which non-linearly realise an internal or space-time group, and which yield a flat or curved scalar manifold. This classification leads to well-known inflationary models such as monomial inflation and $\alpha$-attractors, as well as a new model based on fixed couplings between a dilaton and many axions which non-linearly realises higher-dimensional conformal symmetries. In this model, inflation can be realised along the dilatonic direction, leading to a tensor-to-scalar ratio $r \sim 0.01$ and a spectral index $n_{s} \sim 0.975$. We refer to the new model as ambient inflation since inflation proceeds along an isometry of an anti-de Sitter ambient space-time, which fully determines the kinetic sector.

	\vfill \setcounter{page}{0} \setcounter{footnote}{0}
	\newpage
	
	\tableofcontents
	
\section{Introduction} \label{intro}

Symmetries play a very important role in the construction of effective field theories (EFTs). They offer protection against quantum corrections, can reduce the number of arbitrary coupling constants thereby increasing predictivity and can render small symmetry breaking parameters technically natural. For these reasons, amongst others, gauge and global symmetries often appear in cosmological model building. In this paper, we are interested in non-linearly realised symmetries in the kinetic sector of inflationary models which are weakly broken by an inflation driving potential. 

In the absence of such explicit symmetry breaking, the dynamics of the Goldstone modes is strongly constrained by the non-linearly realised symmetries, resulting in specific signatures in observables. For a scalar field with a shift symmetry $\phi \rightarrow \phi + c$, resulting from the spontaneous breaking of a $U(1)$ internal symmetry, we encounter the Adler zero \cite{adler} with all scattering amplitudes vanishing in the limit where a single external momentum $p$ is taken soft. Generalisations of the Adler zero, where amplitudes vanish more quickly in the soft limit, can be obtained by augmenting the shift symmetry by a space-time extension. Soft amplitudes scaling as $p^{2}$ can be achieved with symmetries which non-linearly realise the five-dimensional Poincar\'{e} group or a non-relativistic contraction (referred to as the Galileon symmetry group). Moreover, there is a unique scalar EFT with a $p^3$ scaling which consists of a specific combination of scalar Galileons \cite{specialgal1,specialgal2}. The interplay between symmetries and soft limits therefore enables one to perform an interesting classification of special EFTs, see \cite{periodic,PSW}.

Most inflationary models, however, include an explicit symmetry breaking term in the form of the scalar potential. The simplest example of such a scenario is realised by single field monomial inflation \cite{chaotic}. Here the scalar's canonical kinetic term is invariant under a shift symmetry which is broken by the potential energy $V = \lambda \phi^{m}$ with integer $m \geqslant 2$, providing a very simple realisation of inflation by a symmetry breaking potential. The symmetry breaking parameter $\lambda$ is constrained to be very small, in Planck units, from the observed level of CMB temperature anisotropies and this is a technically natural scenario. Indeed, this slow-roll choice is not spoiled by field theory or perturbative quantum gravity corrections thanks to the approximate shift symmetry \cite{quantumcorrections} (see chapter 3 of \cite{KLS} for a concise discussion on this topic). 

However, these very simple inflationary models predict large values for the tensor-to-scalar ratio $r$ and have been all but ruled about by CMB polarisation observations \cite{planck,bicep}. This motivates one to investigate slightly more complicated inflationary models which can reduce the value of $r$ without spoiling the radiative stability of the EFT. This is our aim in this paper and we will do so by allowing the scalar potential to break more exotic non-linear symmetries rather than a simple shift. This will require us to construct kinetic sectors with more scalar fields and as we shall see, kinetic sectors which correspond to some non-trivial internal manifold can have interesting observational effects consistent with the current data.

Therefore we will consider models of the form 
\begin{equation} \label{action}
S = \int d^{4}x \sqrt{-g} \left(\tfrac12 M_{\text{pl}}^{2}R - \tfrac12 \Lambda^{4} K(\phi^{I}, \partial_{\mu} \phi^{I}) - V(\phi^{I})\right),
\end{equation}
where $\phi^{I}$ labels $n$ scalar fields, $K(\phi^{I}, \partial_{\mu} \phi^{I})$ is the dimensionless kinetic sector and contains at least two derivatives, and $V(\phi^{I})$ is the symmetry breaking potential. $\Lambda$ is an arbitrary scale introduced on dimensional grounds. Taking a cue from the EFT classification briefly discussed above, we will be interested in cases where the kinetic sector is fixed by a non-linearly realised symmetry corresponding to a coset space $G/H$, where $G$ can be an internal symmetry group or a space-time symmetry group. This will be the main ingredient of the scenarios under investigation; our starting point is conventional in that the 4-dimensional Lorentz group is always linearly realised\footnote{Note that here and in what follows we say that the Lorentz group is linearly realised rather than the full Poincar\'{e} group. In all cases the full Poincar\'{e} group is indeed linearly realised on the fields but translations are non-linearly realised on the space-time coordinates. We adopt this terminology to avoid confusion when we we employ the coset construction.} and our scalar sector is minimally coupled to gravity. As we shall see, coset spaces are a useful way of characterising kinetic sectors for scalar field theories and we note that this has been considered before in the context of inflation in \cite{cliffcosets} and coset spaces have been used to classify condensed matter systems in e.g. \cite{zoo}.

In section 2 we will discuss five different forms for the kinetic sector; three which non-linearly realise an internal symmetry and two which non-linearly realise a space-time symmetry. This exhausts all maximally symmetric possibilities (up to group contractions). For internal symmetries, where each generator commutes with those of the 4-dimensional Poincar\'{e} group, $G$ corresponds to the symmetries of flat, spherical and hyperbolic geometries. For space-time symmetries, where the group $G$ contains the 4-dimensional Poincar\'{e} group, it can correspond to the symmetries of higher-dimensional Minkowski space and anti-de Sitter space\footnote{We do not consider the case where $G$ is the de Sitter group since as far as we are aware this group does not have a 4-dimensional Poincar\'{e} subgroup so it would be impossible to non-linearly realise the de Sitter isometries with scalar fields in 4-dimensional Minkowski space-time (see e.g. \cite{paul} for a discussion of why it is not possible to embed the 4-dimensional Poincar\'{e} group into the 5-dimensional de Sitter group).}. In each case we make use of the coset construction \cite{internal1,internal2,spacetime1,spacetime2} to build invariant kinetic sectors, and will pay most attention to the non-linear realisation of the anti-de Sitter isometries, i.e. the conformal group, since the corresponding coset space has not been well studied in the literature. At this stage let us make it clear that although our scalars are indeed Goldstone bosons, they are not the usual Goldstones appearing in the EFT of inflation \cite{EFTofI1,EFTofI2} since they are not associated with the breaking of time translations. 

We add the symmetry breaking potentials in section 3 where we also couple the scalars minimally to gravity in order to drive inflation. We concentrate most on two examples; one with internal symmetries corresponding to a hyperbolic geometry and the other with space-time symmetries which non-linearly realise the conformal group. For clarity we study the $n=2$ case with two fields, since this captures the main features of the models which are both constructed from a single axion with a shift symmetry and a single dilaton. The former is the $\alpha$-attractors model \cite{alpha-attractors} while the latter we dub ambient inflation, since the non-linear symmetries are those corresponding to a Minkowski 3-brane fluctuating in an anti-de Sitter ambient space-time. 

In both cases we will study the predictions of inflationary trajectories which take place along the dilatonic direction. For a large class of scalar potentials and for order one parameters the predictions of $\alpha$-attractors are in the sweet spot of the Planck data: $n_{s} = 0.968 \pm 0.006$ \cite{planck}. The spectral index takes values close to $0.960$ or $0.967$ depending on our choice of $50$ or $60$ e-folds, while the tensor-to-scalar ratio takes values around $r \sim 0.001$, although it can also be larger. For ambient inflation the spectral index turns out be somewhat bluer than the $\alpha$-attractors prediction taking values between $0.971$ and $0.976$ again for e-folds ranging from $50$ to $60$. There is also a non-trivial difference in the predicted tensor-to-scalar ratio compared to $\alpha$-attractors with ambient inflation naturally predicting $r \sim 0.01$ for order one parameters. Both predictions for the tensor-to-scalar ratio are therefore interesting targets for future ground-based and satellite CMB missions. We present figures in section 3.2 where more accurate values for $n_{s}$ and $r$ are presented for a range of potentials and parameter values. 

Before moving onto the main body of the paper let us first comment on the elephant in the room with regards to large field inflationary models. In comparison to the weak breaking of the shift symmetry in monomial inflation, parameters which break these more exotic symmetries can also be set to a small value in Planck units in a technically natural way. Inflaton loops and graviton loops will not spoil this choice thanks to the weakly broken symmetry. For $\alpha$-attractors this was investigated in \cite{alpha-attractors-corrections} and indeed the hyperbolic geometry of the kinetic sector provides the expected protection. However, a fully fledged theory of quantum gravity is expected to break all continuous global symmetries \cite{globalbreaking} and this phenomenon will manifest itself via symmetry breaking corrections to the inflationary potential of the form $M_{\text{pl}}^{4-n} \phi^{n}$ with order 1 coefficients. For large field inflationary models this can spoil the slow-roll dynamics. Much work has been done to alleviate this problem in the context of monomial inflation i.e. to realise large field inflation while keeping control of these corrections \cite{monodromy1,monodromy2,KS,KLS,inferno} and interestingly these techniques can also be employed to reduce the value of $r$. In terms of axion monodromy models this can be achieved by considering backreaction which can have the effect of flattening the scalar potential \cite{flatten} and in terms of the Kaloper-Sorbo mechanism one can consider field theory corrections and flatten the scalar potential by going into a strongly coupled but controlled regime \cite{london,fourpi}. In this paper our aim is to produce phenomenologically viable theories of inflation which are stable against perturbative quantum gravity effects within EFT. We would therefore require further model building input along these lines to be sure that the potentially troublesome non-perturbative corrections are under control. Indeed this would provide interesting future research directions. 

\section{Kinetic Sector Symmetries} \label{kineticsymms}

In this section we construct five interesting choices for the kinetic sector $K(\phi^{I}, \partial_{\mu} \phi^{I})$ by computing invariant metrics of different coset spaces $G/H$ using the coset construction. 

An important distinction is whether the coset includes the \Poincare group or not. The latter case leads to internal symmetries while the former involves an extension of this space-time symmetry. Internal symmetries are best understood:  the number $n$ of Goldstone bosons  equals $\text{dim}(G/H)$ (i.e.~there is one for every spontaneously broken generator) and their dynamics is uniquely fixed by the non-linearly realised symmetry.

 In contrast, space-time symmetries have the added complication of inverse Higgs constraints \cite{spacetime2} which allow one to build non-linear realisations where the number of Goldstones is less than the number of broken generators: in these cases one has $n \leq \text{dim}(G/H)$ and one can impose inverse Higgs constraints to reduce the non-linear realisation to $n$ scalar fields. We refer the reader to \cite{KRS} for a discussion on the most important aspects of the coset construction for spontaneous breaking of internal and space-time symmetries, and for an introduction to inverse Higgs constraints.  

\subsection{Internal Symmetries}
Firstly, we assume that the non-linearly realised symmetries of the kinetic sector commute with the 4-dimensional Poincar\'{e} group. For maximally symmetric groups, $G$ can be either $ISO(n)$, $SO(n+1)$ or $SO(1,n)$ corresponding to flat, spherical and hyperbolic geometries respectively. Since our aim is to derive kinetic sectors with $n$ scalars we fix $H = SO(n)$ giving us the following three coset spaces
\begin{align}
  \mathbb{R}^n  \simeq  ISO(n) / SO(n)  \,, \qquad
 \mathbb{S}^n  \simeq SO(n+1) / SO(n)  \,, \qquad
 \mathbb{H}^n  \simeq SO(1,n) / SO(n) \,.
\end{align}
In the following we discuss each of these in turn and compute the invariant metrics. For the coset construction we will need the commutators of the Lorentz group in order to compute the Maurer-Cartan form. For $SO(p,q)$ we have
\begin{equation}
[M_{AB},M_{CD}] = \eta_{AC}M_{BD} - \eta_{BC}M_{AD} + \eta_{BD}M_{AC} - \eta_{AD}M_{BC} \,,
\end{equation}
where $\eta_{AB}$ is the metric of a flat space-time with $p$ timelike directions and $q$ spacelike directions. For $ISO(p,q)$ these are augmented with the following nonzero commutators involving the translation generators
\begin{equation}
[M_{AB},P_C] = \eta_{AC}P_{B} - \eta_{BC}P_{A}  \,.
\end{equation}
In all cases one can construct an invariant metric of the form $d \sigma^2 = g_{IJ}(\f)d\f^I d\f^J$ which upon pulling back to 4-dimensional space-time gives rise to
\begin{equation}
g_{IJ}(\f)\partial_\m \f^I \partial_\n \f^J dx^\m dx^\n  \,.
\end{equation}
Since here we are considering non-linearly realised internal symmetries, the $dx^\m$ are invariant under $G$. Therefore the corresponding kinetic sectors are given by
\begin{align}  
K = g_{IJ}(\f)\partial_\m \f^I \partial^\m \f^J \,.
\end{align}
By construction these are invariant under the linearly realised Lorentz group as well as the non-linearly realised isometry group of the coset. A related discussion of these cosets can be found in e.g.~\cite{cliffcosets} (in particular the case $n=2$).

\subsubsection*{Flat Geometry}
For the first coset space the translations of $ISO(n)$ are broken while the rotations are unbroken. If we let $i$ be an $SO(n)$ index then the only $H$-invariant way of parametrising the coset space is
\begin{equation}
\gamma = e^{\phi^{i}P_{i}} \,,
\end{equation}
where $\phi^{i}$ are the Goldstone bosons. The Maurer-Cartan form is very simple to calculate in this case since $[P_{i},P_{j}] = 0$ and is given by
\begin{equation}
\gamma^{-1} d \gamma = d \phi^{i} P_{i}.
\end{equation}
It follows that the only $SO(n)$ invariant metric one can construct is 
\begin{equation}
d \sigma^{2} = \delta_{ij} d \phi^{i} d \phi^{j},
\end{equation}
corresponding to a flat scalar manifold. The kinetic sector therefore reads
 \begin{align}
  K = \delta_{ij} \partial_{\mu} \phi^i \partial^{\mu} \phi^j \,,
 \end{align}
where the Goldstones have the dimension of length such that $K$ is dimensionless. Each Goldstone inherits a shift symmetry $\phi^{i} \rightarrow \phi^{i} + c^{i}$ from the spontaneously broken translations. Arbitrary functions of this two-derivative combination (similar to $P(X)$ theories for a single scalar) will also be invariant, but such higher-order corrections will not play a role in our application to slow-roll inflationary models.

\subsubsection*{Spherical Geometry}

For a spherical geometry we have $G = SO(1+n)$ and we denote the generators of this group as $M_{1i}$ and $M_{ij}$ where again $i,j$ are $SO(n)$ indices. In this case, the broken generators are $M_{1i}$ and the only $H$-invariant parametrisation of the coset element is 
\begin{equation}
\gamma = e^{\phi^{i}M_{1i}}.
\end{equation}
The resulting Maurer-Cartan form is not very illuminating but leads to the following unique choice for a $G$-invariant metric
\begin{equation}
d \sigma^{2}  = \frac{\sin^{2} \sqrt{\phi^{2}}}{\phi^{2}} \delta_{ij} d \phi^{i} d \phi^{j} + \left( 1 - \frac{\sin^{2} \sqrt{\phi^{2}}}{\phi^{2}} \right) \frac{\phi_{i} \phi_{j}}{\phi^{2}} d \phi^{i} d \phi^{j} ,
\end{equation}
where $\phi^{2} = \delta_{ij} \phi^{i} \phi^{j}$. Note that the metric is manifestly invariant under the linearly realised $SO(n)$, and can be used to construct the corresponding kinetic term. Restricting to $n=2$ for simplicity, a more familiar form might be
\begin{equation}
d \sigma^{2} = L^{2} (d \theta^{2} + \sin^{2} \theta d \varphi^{2}) = 4 L^{4} \frac{dZ d \bar{Z}}{(L^{2} + |Z|^{2})^{2}} \,,
\end{equation}
where $Z =\f + i \pi = L e^{i \varphi} \tan \theta / 2 $ with $0 \leqslant  \theta < \pi$ and $0 \leqslant  \varphi < 2\pi$.  The corresponding kinetic sector is therefore
 \begin{equation} 
  K = \frac{4L^4}{(L^2 + \f^2 + \pi^2)^2}\big((\partial \phi)^{2} + (\partial \pi)^{2}\big) \,,
 \end{equation} 
where we have explicitly included the length scale $L$ which sets the radius of the sphere.

\subsubsection*{Hyperbolic Geometry}
For a hyperbolic geometry we have $G = SO(1,n)$ and we denote the generators of this group as $M_{0i}$ and $M_{ij}$ where again $i,j$ are $SO(n)$ indices. In this case the broken generators are $M_{0i}$ and the only $H$-invariant parametrisation of the coset element is 
\begin{equation}
\gamma = e^{\phi^{i}M_{0i}}.
\end{equation}
As with the spherical case the Maurer-Cartan form is somewhat complicated, but one can easily show that it leads to the following unique choice for a $G$-invariant metric
\begin{equation}
d \sigma^{2}  = \frac{\sinh^{2} \sqrt{\phi^{2}}}{\phi^{2}} \delta_{ij} d \phi^{i} d \phi^{j} + \left( 1 - \frac{\sinh^{2} \sqrt{\phi^{2}}}{\phi^{2}} \right) \frac{\phi_{i} \phi_{j}}{\phi^{2}} d \phi^{i} d \phi^{j} \,,
\end{equation}
where again $\phi^{2} = \delta_{ij} \phi^{i} \phi^{j}$. For $n=2$ we can write this metric as 
\begin{equation}
d \sigma^{2} = L^{2}(d \tau^{2} + \sinh^{2} \tau d \theta^{2}) = 4 L^{4} \frac{dZ d \bar{Z}}{(L^{2} - |Z|^{2})^{2}} \,, \label{disc}
\end{equation}
where now $Z = \f + i\pi = L e^{i \theta} \tanh \tau / 2$ with $0 \leqslant  \tau < \infty$ and $0 \leqslant  \theta < 2\pi$, and the corresponding kinetic sector is
\begin{equation} 
K =  \frac{4 L^4}{(L^2 - \f^2 - \pi^2)^2}\big((\partial \phi)^{2} + (\partial \pi)^{2}\big) \,,
\end{equation}  
where now $L$ sets the curvature radius of the hyperbolic geometry. 

\subsection{Space-time Symmetries}
Now we consider non-linearly realised symmetries which do not commute with the 4-dimensional Poincar\'{e} group. If we again assume maximal symmetries then in principle we would have three possibilities for $G$ corresponding to the isometries of higher-dimensional Minkowski space, de Sitter space and anti-de Sitter space. However, since there is no way to embed the 4-dimensional Poincar\'{e} group into the de Sitter group we only have the two remaining possibilities. In each case we take $H = SO(1,3) \times SO(p)$, leading to  the following two coset spaces 
\begin{eqnarray}
&& \text{Mink}_{4+n}: ISO(1,3+n) / (SO(1,3) \times SO(n)) \,, \notag \\
&& \text{AdS}_{4+n}: SO(2,3+n)/ (SO(1,3) \times SO(n-1)) \label{adscoset}. 
\end{eqnarray}
All of these yield non-linear realisations constructed from $n$ scalars, where one has to impose inverse Higgs constraints to remove the additional so-called inessential Goldstone modes.

Since the non-linearly realised symmetries no longer commute with the 4-dimensional Poincar\'{e} group, the $dx^\m$ are not invariant and one cannot construct invariant kinetic sectors in the same way as for the internal case. Rather, the invariant that is lowest order in derivatives is a Poincar\'{e} invariant combination of four copies of the Maurer-Cartan  components $e^\m{}_\n dx^\n$ (corresponding to the translation generators $P_\m$) in the following way
\begin{equation}
\epsilon_{\m\n\r\s} (e^\m{}_{\a} dx^\a) \wedge (e^\n{}_\b dx^\b) \wedge (e^\r{}_\g dx^\g) \wedge   (e^\s{}_\delta  dx^\delta) \,.
\end{equation} 
As we will see, the above term is not always strictly a kinetic term with at least two derivatives. In some cases there is also a potential term necessary to ensure invariance. 
 
We will now briefly review the flat case, and then discuss the AdS$_{4+n}$ coset space in more detail, since to our knowledge the resulting invariants have not been constructed before for general $n$. 

\subsubsection*{Minkowski Space}

For the Minkowski coset space, the broken generators are translations $P_{i}$ and Lorentz transformations $M_{\mu i}$ where $i$ is an $SO(n)$ index and $\mu$ is a 4-dimensional space-time index. Given the intricate link between the coset parametrisation and one's ability to impose inverse Higgs constraints \cite{KRS}, we parametrise the coset element as
\begin{equation}
\gamma = e^{x^{\mu}P_{\mu}} e^{\phi^{i} P_{i}} e^{\Omega^{\mu i } M_{\mu i }} \,,
\end{equation}
from which we can compute the Maurer-Cartan form using the commutators of the $(4+n)$-dimensional Poincar\'{e} group. Note that for space-time symmetry breaking we include the unbroken translations $P_{\mu}$ in the coset element since they act non-linearly on the space-time coordinates. Again the full form of the Maurer-Cartan form is not particularly useful but after we impose inverse Higgs constraints to remove the vectors $\Omega^{\mu i}$ the kinetic sector is seen to be \cite{wess}
\begin{equation}
K = \sqrt{- \text{det}(\eta_{\mu\nu} + \partial_{\mu} \phi^{i} \partial_{\nu} \phi_{i})} \,.
\end{equation}
One can also derive this term by calculating the induced metric corresponding to a Minkowski 3-brane embedded in higher-dimensional Minkowski space \cite{higher}.  The kinetic sector then simply corresponds to the measure of this metric. Similarly, invariants like the Einstein-Hilbert term give rise to higher-order invariants of this symmetry\footnote{In the single field case, these are so-called DBI galileons \cite{reunited} which have interesting behaviour in their soft amplitudes due to the non-linearly realised symmetry \cite{specialgal1,PSW}, as mentioned before.}.

Note that for space-time symmetries we need to include a whole tower of operators to non-linearly realise the broken symmetry group in the kinetic sector, whereas for the internal symmetries discussed above this could be achieved order by order in derivatives. This crucial difference follows from the transformation properties of the metrics derived from the coset construction: they are invariant for the internal cases while they transform covariantly for the space-time symmetry cases.
 
\subsubsection*{Anti-de Sitter Space}
The final coset we consider corresponds to the spontaneous breaking of the anti-de Sitter isometries, corresponding to the coset space \eqref{adscoset}.
To derive our kinetic sector we will make use of the so-called AdS basis for the conformal algebra defined by the following non-vanishing commutators\footnote{Note that we have set the AdS radius $L = 1/\sqrt{2}$ but will reintroduce it later on.} \cite{equivalence} 
\begin{equation*}
\begin{aligned}[c]
&[P_A,D] = P_A \, \\
&[\hat{K}_A,D] = - \hat{K}_A  + P_A \,\\
&[P_A,\hat{K}_B] = 2 M_{AB}  + 2\eta_{AB}D\, \\
&[\hat{K}_A,\hat{K}_B] = 2M_{AB} \,
\end{aligned}
\qquad
\begin{aligned}[c]
&[M_{AB},P_C] = \eta_{AC}P_B - \eta_{BC}P_A \\
&[M_{AB},\hat{K}_C] = \eta_{AC}\hat{K}_B - \eta_{BC}\hat{K}_A \\
&[M_{AB},M_{CD}] = \eta_{AC}M_{BD} - \eta_{BC}M_{AD} + \eta_{BD}M_{AC} - \eta_{AD}M_{BC}
\end{aligned}
\end{equation*}
where $A = 0,1,\ldots 2+n$. The relation to the standard basis is given by
\begin{equation}
\hat{K}_{A} = K_{A} + \frac{1}{2}P_{A}.
\end{equation}
Here, in addition to the Poincar\'{e} generators, we have $D$ and $K_{A}$ which are the generators of dilatations and special conformal transformations respectively. The AdS basis is useful since the resulting kinetic sector matches the one from embedding Minkowski 3-branes in $(4+n)$-dimensional anti-de Sitter space\footnote{We refer the reader to \cite{KRS,mapping} for discussions on the physical equivalence of the resulting non-linear realisations with the conformal group defined in these two different bases.}.

Again given the discussion in \cite{KRS}, we parametrise the coset element as
\begin{equation} \label{adsconformalcoset1}
\g  = e^{x^{\m}P_{\m}}e^{\pi^{i}P_{i}} e^{\varphi D} e^{\Omega^{\m i}M_{\m i}}  e^{\psi^{\m}\hat{K}_{\m}}e^{\sigma^{i}{\hat{K}_{i}}} \,,
\end{equation}
where we have assumed $n >1$ and $\mu = 0,\ldots 3$ and $i = 4 \ldots 2 +n$. Non-linear realisations of this symmetry breaking can then be constructed from the Maurer-Cartan form which can be written as
\begin{align} 
\g^{-1}d\g = \o^\m P_\m + \o^A T_A + \o^i T_i \,,
\end{align} 
where $T_{A}$ are the broken generators of the conformal group and $T_{i}$ are the unbroken ones i.e. $M_{\mu\nu}$ and $M_{ij}$. To match our previous notation we have $\o^\m = e^{\mu}{}_{\nu} dx^{\nu}$ once we pull back to 4-dimensional space-time. The commutators
\begin{eqnarray}
[P_{\mu},M_{\nu i}] \supset \eta_{\mu\nu} P_{i} \,, \qquad
[P_{\mu},\hat{K}_{\nu}]\supset \eta_{\mu\nu}D \,, \qquad
[P_{\mu},\hat{K}_{i}] \supset M_{\mu i}  \,,
\end{eqnarray}
ensure that the fields $\Omega^{\m i}$, $\psi^{\m}$ and $\sigma^{i}$ appear linearly in the Maurer-Cartan components along the broken generators $P_{i}$, $D$ and $M_{\mu i}$ respectively and given our choice for the coset parametrisation they only appear algebraically. We can therefore impose inverse Higgs constraints to eliminate all of these fields in favour of $\varphi$, $\pi^{i}$ and their derivatives leaving us with a non-linear realisation constructed from the dilaton and $n-1$ axions. 

The axion fields $\pi^{i}$ are guaranteed to be shift symmetric since they are the Goldstones of broken translations and they will inherit a linearly realised $SO(n-1)$ symmetry due to the unbroken rotations $M_{ij}$. Without loss of generality we can therefore simply consider the case where $n=2$ i.e. where there is a single axion $\pi^{4} = \pi$ then augment the resulting non-linear realisation by adding the other axions in an $SO(n-1)$ invariant manner. We therefore consider the coset space and coset element\footnote{Here we have $\Omega^{\mu 4} = \Omega^{\mu}$ and $\sigma^{4} = \sigma$.}
\begin{equation} 
SO(2,5) / SO(3,1)  \,, \qquad 
\g  = e^{x^{\m}P_{\m}}e^{\pi P_{4}} e^{\varphi D} e^{\Omega^{\m }M_{\m 4}}  e^{\psi^{\m}\hat{K}_{\m}}e^{\sigma {\hat{K}_{4}}} \,,
\end{equation}
for concreteness.

As with all of the previous cases our aim is to compute a metric from which we can derive $G$-invariant theories. From the coset construction the metric is fixed in terms of $\omega^{\mu}$ so we only need to compute this contribution to the Maurer-Cartan form and the necessary inverse Higgs constraints. A somewhat lengthy calculation yields the following kinetic sector\footnote{Here we have made the rescaling $\varphi \rightarrow \sqrt{2} \varphi$ and reintroduced the AdS radius $L$.}
\begin{equation}
K =  \sqrt{-\text{det}\left(e^{2 \varphi / L} (\eta_{\mu\nu} + \partial_{\mu} \pi \partial_{\nu} \pi) +  \partial_{\mu} \varphi \partial_{\nu} \varphi \right)} ,
\end{equation}
and by adding the remaining axions we arrive at 
\begin{equation}
K =  \sqrt{-\text{det}\left(e^{2 \varphi / L} (\eta_{\mu\nu} + \partial_{\mu} \pi^{i} \partial_{\nu} \pi_{i}) +  \partial_{\mu} \varphi \partial_{\nu} \varphi \right)}.
\end{equation}
Again a full tower of operators is required to non-linearly realise the conformal symmetries.

As we mentioned above this is precisely what one gets from computing the world-volume of a Minkowski 3-brane embedded in $(4+n)$-dimensional AdS space. This is easy to see with the AdS$_{4+n}$ metric written in the Poincar\'{e} patch
\begin{equation}
d\sigma^{2} = \frac{L^{2}}{z^{2}} (\eta_{\mu\nu} dx^{\mu}dx^{\nu} + dy_{i}dy^{i} + dz^{2}) \,,
\end{equation}
where $L$ is the AdS radius, $x^{\mu}$ are the 4-dimensional brane directions, $y^{i}$ are the axionic directions and $z$ is the dilatonic direction. In fact this metric motivates us to make the field redefinition $\phi =L  e^{-\varphi/L}$ such that the kinetic sector becomes\footnote{In the single field case this is the kinetic sector of DBI inflation \cite{dbiinflation} up to field redefinitions. Note that only in the single field case can we canonically normalise.} 
\begin{equation}
 K =  \sqrt{-\text{det}\left(\frac{L^{2}}{\phi^{2}} \left(\eta_{\mu\nu} + \partial_{\mu} \pi^{i} \partial_{\nu} \pi_{i} +  \partial_{\mu} \phi \partial_{\nu} \phi\right)\right)}. \label{AdS-kinetic}
\end{equation}
We remind the reader that $K$ is dimensionless and the Goldstones $\phi$ and $\pi^{i}$ have dimension of length.

In what follows we will use this kinetic structure, which is protected by the non-linearly realised conformal symmetry, to realise slow-roll inflation. When we expand the square root we see that there is already a potential of the form $(L / \phi)^4$, which prevents us from interpreting this as a purely kinetic term. In the case where $n=1$, corresponding to $\pi^{i} = 0$, we know that there is a Wess-Zumino term which one can add to the action to remove this potential without breaking the symmetries; see \cite{reunited} for a discussion of this term in the context of embedded branes and \cite{wess} for a derivation using the corresponding Maurer-Cartan form. In the following we shall assume that there is a Wess-Zumino term for arbitrary $n$ which we can add to the action which reduces to the $n=1$ case when we send $\pi^{i} = 0$. We simply denote this Wess-Zumino as W.Z such that our symmetric kinetic sector is 
\begin{equation} 
K =   \frac{ L^{4}}{\phi^{4}} \sqrt{-\text{det}(\eta_{\mu\nu} +  \partial_{\mu} \pi^{i} \partial_{\nu} \pi_{i} + \partial_{\mu}\phi \partial_{\nu}\phi)} + \text{W.Z}  \,,  \label{AdS-kinetic-WZ}
\end{equation}
which has a tower of higher-derivative operators in order to realise the non-linear AdS$_{4+n}$ symmetry. Some of the non-linear AdS$_{4+n}$ symmetries are realised order by order in derivatives, namely, the shifts in $\pi^{i}$ associated to the translation generators $P_{i}$, and the dilations $D$. In contrast, the transformations associated with the broken Lorentz generators $M_{\mu i}$ and the special conformal transformations $K_{\mu}$ and $K_{i}$ require the full tower. The precise from of these final symmetries is quite lengthy so we don't present them here but one can easily extract them along the lines discussed in \cite{higher} by using the isometries of the AdS$_{4+n}$ space and the appropriate embedding functions of the Minkowski 3-brane. This way of computing the symmetries is more straight forward than via the coset construction.   

\subsection{Geometrical Considerations} \label{geometricalconsiderations}

In this paper we are interested in slow-roll inflationary applications of the above kinetic sectors. Under this approximation, we can neglect higher-order terms in derivatives. This yields a two-derivative kinetic sector which is amenable to a geometric interpretation. For simplicity we will discuss the two-field case in this section (while the generalisation to more scalars is rather straightforward). We therefore have different geometries on the complex plane, with three interesting geometric possibilities, in addition to the flat case. 

There are two possibilities with negative curvature which are interesting to compare against each other. In particular, the hyperbolic kinetic sector is based on the geometry \eqref{disc} in terms of disc coordinates. An alternative parametrisation is in terms of half-plane coordinates $T$, which are related via the Cayley transformation
\begin{align}
 \frac{T}{L} =  \frac{L-Z}{L+Z} \,, \label{Cayley}
\end{align}
where we keep all coordinates as lengths and in the following drop all order one factors since they can always be absorbed into a redefintion of $L$. This brings the hyperbolic geometry to the form
 \begin{align}  
  d \sigma^2 = \frac{L^{2}}{(T+ \bar T)^2} dT d \bar T \,. \label{pole-two}
 \end{align} 
In contrast, the truncation of the square root structure in the AdS kinetic sector \eqref{AdS-kinetic-WZ} at two derivative order naturally leads us to consider the geometry\footnote{A more detailed knowledge of the structure of the Wess-Zumino term is required to be sure that at the two derivative level the AdS kinetic sector gives rise to the geometry (\ref{pole-four}). In this paper our ultimate interest is in inflationary dynamics along the dilatonic direction where this subtlety plays no role, but here we will point out the geometric properties at the two derivative level assuming that the Wess-Zumino allows for a truncation to (\ref{pole-four}).}
 \begin{align}
   d \sigma^2 = \frac{L^{4}}{(T+ \bar T)^4} dT d \bar T \,. \label{pole-four}
 \end{align}
Both of these lead to an axion-dilaton system, with the dilaton being the real part of $T$ and the shift symmetric axion being the imaginary part i.e. $T = \phi + i \pi$. Importantly, the couplings between the two scalars is different in both cases, dictated either by an internal or space-time symmetry. Again, this crucial difference follows from the invariance or covariance of the associated coset metric.
 
The two relevant geometries are therefore special cases of the more general metric
 \begin{align} \label{generalmetric}
   d \sigma^2 = \frac{L^{p}}{(T+ \bar T)^p} dT d \bar T \,,
 \end{align}
for an arbitrary power $p$, which we will refer to as the order of the pole. The above discussion singles out three values, $p=0, 2$ and $4$, as being special from a symmetry perspective. The former two have an enhanced isometry group at the two-derivative level while the latter realises the full symmetries of the AdS$_6$ group once the higher order corrections are taken into account. 

Of the maximally symmetric possibilities, $p=0$ is flat and has an $ISO(2)$ isometry group. The case $p=2$ is the hyperbolic half-plane and has the Mobius transformations as isometry group, isomorphic to $SO(2,1)$. This ensures that the curvature of the manifold is constant and negative with scaling $R \sim  L^{-2}$. Moreover, the Mobius transformations include the inversion $T \rightarrow 1/T$. As a consequence, there is a pole of order two at $T=0$ as well as at $T=\infty$. Indeed, one can see that the proper distance to both points is infinite. For a geometry of the form (\ref{generalmetric}) $p=0$ and $p=2$ are the only ones with maximal symmetries and no singularities. 

For any other value of $p$ there is only a single isometry: the shift in the axionic direction. However, as we have discussed, $p=4$ is special for other reasons since when we include the higher order operators we can realise the full AdS$_{6}$ symmetries. At the two derivative level, of these symmetries the linearly realised Lorentz group of course survives the truncation to the two-derivative level but so does the shift in $\pi$, $\pi \rightarrow \pi +c$, corresponding to the broken translation $P_{4}$ and the scale symmetry. For the field basis used in \eqref{AdS-kinetic-WZ} this symmetry reads
\begin{equation}
\phi \rightarrow \phi + \lambda(\phi - x^{\mu} \partial_{\mu} \phi) \hspace{1cm} \pi \rightarrow \pi + \lambda(\pi - x^{\mu} \partial_{\mu} \pi) ,
\end{equation}
where $\lambda$ is the infinitesimal parameter of dilatations and here we do not transform the space-time coordinates. The other symmetries corresponding to the broken Lorentz transformations and special conformal transformations require higher order operators for invariance of the action. In the following we will investigate the effects of these symmetries on inflationary dynamics but we note that spontaneous breaking of scale invariance has been studied in the context of inflation before, e.g. \cite{nemanjascale}, but in a different set-up to ours. 

For $p>2$ the proper distance to the pole at $T=0$ is infinite but this does not hold for the point $T = \infty$. Given that in these cases the curvature scales as $R \sim (T + \bar{T})^{p-2}$, the geometry has a singularity at this point. Note that $p$ and $4-p$ yield identical results along the real line but this is not true for the entire complex plane. We can use the Cayley transformation \eqref{Cayley} to go to disc-like coordinates, which highlights both special points at $T=0$ and $T=\infty$ and moves them to finite coordinate values. This leads to the space-time interval
 \begin{align}
  d \sigma^2 =   \frac{ L^{4} (L+Z)^{p-2} (L+ \bar Z)^{p-2}}{(L^{2}-Z \bar Z)^p} dZ d \bar Z \,.
 \end{align}
For $p=2$, this choice of coordinates highlights the $SO(2)$ isometry and corresponds to the Poincar\'{e} disc parametrisation of the hyperbolic geometry. Along the real axis, the above metric reduces to the interval
 \begin{align}
  d \sigma^2 =  \frac{ L^{4} (L+Z)^{p-4}}{(L-Z)^p} dZ^2 \,.
 \end{align}
For $p=2$, there is equivalence between the two special points at the real line, $Z = \pm 1$. For values $p>2$ we have mapped the pole to $Z=1$ and the singularity to $Z=-1$, while these two are interchanged for $p<2$. Again $p=0$ and $p=4$ are special in that there is no singularity at one side. We refer the reader to \cite{alphageometry} for a more detailed discussion for $p=2$ and to \cite{rubio} for a further discussion on the role of geometry in scale invariant models of inflation.

\section {Symmetry Breaking Potentials} \label{addingpotential}

\subsection{Universality Classes of Inflation}
The symmetric kinetic sectors of the previous section provide an attractive starting point for inflationary scenarios. To this end, one has to introduce a scalar potential in order to introduce the required energy for the accelerated expansion, as well as evolution towards the end of inflation. At the same time, the weakly broken symmetry in the kinetic sector protects the model against large quantum corrections within EFT. This ties in with the smallness of the observed level of quantum fluctuations: the inflationary energy scale is orders of magnitude below the Planck scale and small symmetry breaking parameters are technically natural.

As alluded to in the introduction, the simplest of such examples consists of a single scalar field with a canonical kinetic term. The symmetries of this model include a constant shift which will be broken by the introduction of a generic potential. We will assume a Minkowski minimum somewhere in field space, which can be taken at $\phi=0$ without loss of generality. Different manners of breaking this shift symmetry then correspond to e.g.~a quadratic or a quartic scalar potential around this point, or a combination of such monomials.

The inflationary predictions of such models are particularly simple under the assumption that a single monomial dominates the inflaton dynamics at the observable window of $N=50$ to $60$ e-folds. Taking $V=\lambda \phi^m$ as the simplest example of this class, the resulting predictions are 
 \begin{align}
  n_s = 1 - \frac{2+m}{2} \frac{1}{N} \,, \qquad r = \frac{4m}{N} \,. \label{mono-predictions}
 \end{align}
The leading order $1/N$ scaling for the tensor-to-scalar ratio means that the simplest of these models with e.g. $m=2$ or $m=4$ are virtually ruled out \cite{planck,bicep}. 

The underlying assumption in the above is that $N=50$ to $60$ e-folds is a generic window on the primordial quantum fluctuations, i.e.~there is nothing special about the moment we probe inflation, leading to an expansion in the small parameter $1/N$. Different models (e.g.~of polynomial character) will lead to the same predictions at leading order in $1/N$ (under the assumption that $N=50$ to $60$ is dominated by a single monomial) and only lead to different, model-dependent subleading terms at higher order in $1/N$. 

However, the above predictions \eqref{mono-predictions} are one out of two possible perturbative expansions\footnote{A third possibility has a non-pertubative expansion in $1/N$ instead \cite{Garcia}, which includes natural inflation \cite{Freese}. These can be associated to the kinetic sector with the positively curved internal manifold of section 2.} in $1/N$. The alternative has predictions that can be written as (with $p>1$) \cite{universality}
\begin{eqnarray}
n_{s} = 1 - \frac{p}{p-1} \frac{1}{N} \,, \qquad r = \frac{r_0}{N^{p/(p-1)}} \,, \label{pole-predictions}
\end{eqnarray}
at leading order in the $1/N$ expansion. Instead of a monomial expansion of the scalar potential, the second class of inflationary predictions can be conveniently parametrised in terms of the kinetic sector and can result in a suppression of tensor modes. Rather than having a canonical kinetic term, one can allow for a pole in the kinetic sector of the theory. The latter is a natural possibility in multi-field inflation, as suggested by UV theories, where in general one cannot canonically normalise the fields. Along the single-field trajectory, the kinetic sector has the general Laurent expansion \cite{unity}, see also \cite{pole, contour},
\begin{equation} 
 K =  \left( \frac{a_p}{\phi^{p}} + \frac{a_{p-1}}{\phi^{p-1}} + \ldots \right) (\partial \phi)^{2} \,,
\end{equation}
where $p$ is the order of the pole. The assumption in this scenario is that $V$ is regular around $\phi = 0$ i.e. we have 
\begin{equation}
V = V_{0}(1 + c_1 \phi + c_2 \phi^2 + \ldots).
\end{equation}
As inflation takes place close to the pole, it is only the leading term in the scalar potential which determines the inflationary predictions and pole inflation can therefore be seen as a very convenient parametrisation of these inflationary models: all relevant information about the prediction is stored in the leading term of the kinetic sector. Note that for $p=2$ the coefficient $c_1$ drops out of all observables due to the scaling symmetry in the kinetic sector, while for other values it can always be set equal to unity by a rescaling of the field and redefinitions of $a_{p}, a_{p-1}$ etc. Many models fall in the same universality class, with the same model-independent leading predictions and different model-dependent sub-leading terms at higher order in $1/N$.

Our previous discusssion has singled out two types of poles that have an enhanced symmetry in the kinetic sector. The first is the well-known inflationary model based on the hyperbolic geometry \eqref{pole-two} with $p=2$ which is commonly referred to as $\alpha$-attractors \cite{Porrati, alpha-attractors}. In this case, under the assumption of a regular potential at the pole, the inflationary predictions yield
\begin{eqnarray}
 n_{s} = 1- \frac{2}{N} \,, \qquad r = \frac{8 a_{2}}{N^{2}} \,, \label{alpha-predictions}
\end{eqnarray}
to leading order in $1/N$ where we have set $M_{\text{pl}} = \Lambda = 1$ and the higher order corrections are sufficiently suppressed \cite{universality}. As discussed in the literature, the leading terms in the $1/N$ expansion are model-independent for $\alpha$-attractors, with robust predictions. The fixed $2/N$ deviation from scale invariance is in very good agreement with Planck constraints\cite{planck}, leading to e.g.~$n_s=0.960$ to $0.967$ for a range of $N=50$ to $60$ (as will be the case for all following quotes). Moreover, the $1/N^2$ scaling implies that $r$ generically takes values of a few permille, assuming order one values for $a_{2} = 3 \alpha/2$. The benchmark model $a_{2}=3/2$, which corresponds to Starobinsky, has $r=0.005$ to $0.003$, while other constructions can boost this to percent level values, see e.g.~\cite{Ferrara:2016, Wrase}.

The other inflationary model based on the AdS kinetic sector, which has not been discussed previously in the literature, corresponds to pole inflation with $p=4$ as shown in \eqref{pole-four}, plus higher order corrections in the kinetic sector which will be suppressed during inflation. Inflation proceeds along one of the isometries of the AdS space: as inflation proceeds, the 3-brane moves from through the ambient space. It is therefore natural to refer to this set-up as ambient inflation. It follows from the above discussion that the inflationary predictions to leading order in $1/N$ are 
\begin{eqnarray}
n_{s} =  1 - \frac{4}{3 N} \,, \qquad r = \frac{8 a_{4}^{1/3}}{3^{4/3}}\frac{1}{N^{4/3}} \,, \label{ambient-predictions}
\end{eqnarray}
where again we have set $M_{\text{pl}} = \Lambda = 1$. For order one values of $a_{4}$ this leads to a spectral index with a range $n_s = 0.973$ to $0.978$ which is compatible with observational constraints if the number of e-folds would be on the low side of the range from $50$ to $60$ \cite{planck}. However, it turns out that the next to leading order correction to $n_{s}$ scales as $1 / N^{4/3}$ and given the sensitivity of CMB experiments this can produce important corrections. Fortunately this next correction comes in with a minus sign so it can decrease the value of $n_{s}$ thereby moving it towards the sweet spot of the Planck data. We will discuss this further in subsection 3.2. The $1/N^{4/3}$ scaling for the tensor-to-scalar ratio naturally takes values at the percent level, for instance $r = 0.010$ to $0.008$ for $a_{4} = 1$. The tensors therefore come out an order of magnitude higher than the generic $\alpha$-attractor prediction. Both models therefore provide interesting observational targets for upcoming ground-based (e.g.~CMB-S4) and satellite (e.g.~Litebird and Core) CMB polarisation experiments.

For $p >1$ the number of inflationary e-folds is given by
\begin{equation}
N = \int \frac{a_{p}}{ \phi^{p}} d \phi \sim \frac{a_{p} \phi^{1-p}}{(p-1)} ,
\end{equation}
from which we can extract the field range during inflation. For $\alpha$-attractors ($p=2$) we have $N \sim \phi^{-1}$. After we canonically normalise the kinetic sector we see that the field range scales as $\sim \text{log}(N)$ in Planck units. On the other hand, for ambient inflation ($p=4$) the field range of the canonically normalised field scales as $\sim N^{1/3}$ in Planck units. Both of these field ranges are smaller than for chaotic inflation where the field range scales as $\sim N^{1/2}$. However, note that both pole models are essentially large-field models with super-Planckian excursions. They could therefore be susceptible to UV considerations such as the weak gravity conjecture, see e.g.\cite{WGC1,WGC2,WGC3}, which questions the validity of the EFT but this question is far from settled and remains an area of active research.

Even in the context of EFT, in the above we have assumed that inflation takes place along the dilatonic direction, while setting the axion constant. For $\alpha$-attractors the resulting predictions turn out to be a very good approximation to those where the axion is not stabilised thanks to the hyperbolic geometry of the scalar manifold \cite{doublealpha}. In the absence of some mechanism to stabilise the axion in ambient inflation we would require a similar analysis to be sure that our predictions are stable against turning on axionic fluctuations. Again in ambient inflation the kinetic sector is fixed by symmetry so there is reason to believe that this will indeed be the case but the reader should bear in mind that a full multi-field analysis would require a more detailed knowledge of the form of the Wess-Zumino term in the AdS kinetic sector (\ref{AdS-kinetic-WZ}).

\subsection{Adding Curvature to Reduce Tensors} \label{lowerr}

A natural question concerns the relation between the different cases above: is there a limit in which the curved cases, with poles of order $2$ and $4$, reduce to the flat case with $p=0$ and the monomial expansion? We will now show that this is indeed the case. The simplest way to this connection is in terms of the $Z$ coordinates introduced in subsection \ref{geometricalconsiderations}. Along the real line, this yields kinetic sectors
 \begin{align}
  K = \frac{1}{(1-\phi^2/L^2)^2} (\partial \phi)^2 \,, \qquad K = \frac{1}{(1-\phi/L)^4} (\partial \phi)^2 \,, \label{kinetic-sectors}
 \end{align}
for $p=2$ and $p=4$, respectively. In this section we will again work with $M_{\text{pl}} = \Lambda = 1$. In this parametrisation, the kinetic structure is regular around the minimum at $\phi=0$ while the pole is located at $\phi=L$. One can thus go from the flat case, with $L$ infinite, to the curved case by bringing the pole in from infinity to a finite distance from the minimum of the potential. This allows for a continuous interpolation between the different flat and curved inflationary scenarios of the previous subsection.

\begin{figure}[b!]
\centering
\includegraphics[scale=.55]{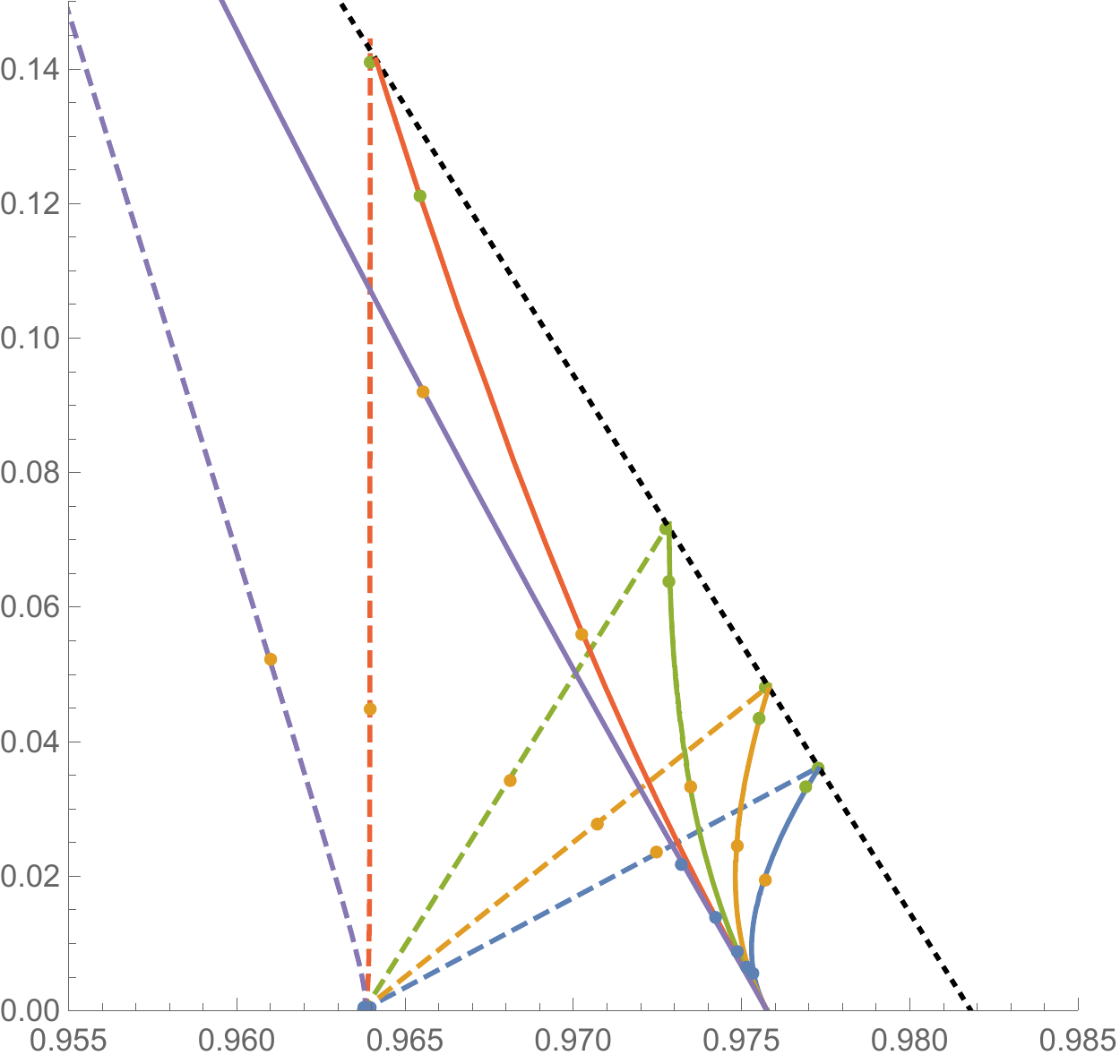} \;\;\;
\includegraphics[scale=.55]{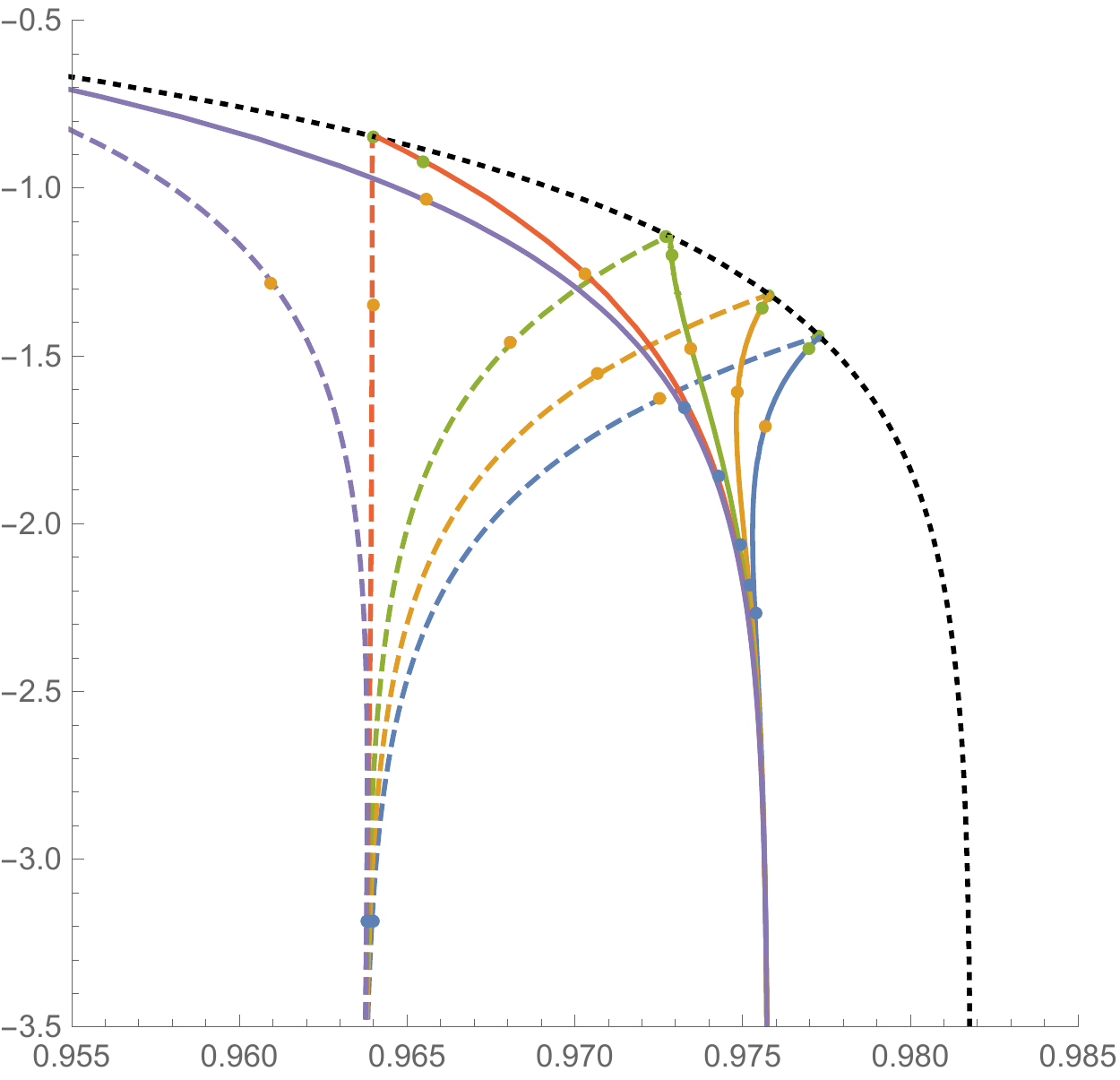}
\caption{\it \small{The interpolation in $(n_s,r)$ (in linear and log plots) from the flat predictions \eqref{mono-predictions} to the $\alpha$-attractor predictions \eqref{alphaplot} based on an internal hyperbolic symmetry (dashed lines) or to the ambient inflation predictions \eqref{adsplot} based on a spacetime AdS symmetry (solid lines).  The different lines indicate models with a monomial potential $\phi^m$ with $m=(4,2,1,2/3,1/2)$ from left to right, while the curvature scale $L$ in the kinetic sectors \eqref{kinetic-sectors} varies along each line, with blue, orange and green dots at $L=1,10,100$ respectively. The black dotted line indicates the flat limit $L \rightarrow \infty$ with canonical kinetic term and monomial potentials. We have taken $N=55$ throughout.}}
\label{fig:ClassI}
\vspace{-.3cm}
\end{figure} 

We have illustrated the behaviour of the different inflationary predictions in the presence of a symmetry breaking potential $V = \lambda \phi^{m}$ in the plots below. Included are the predictions based on the flat geometry with a weakly broken shift symmetry, as well as the two geometries with negative curvatures. 
\begin{itemize}
\item The inflationary predictions based on the flat geometry include those of monomial inflationary models, with a $1/N$ scaling for both the spectral index as well as the tensor-to-scalar ratio, indicated by the dotted line.
\end{itemize}
Turning on the curvature scale $L$ of the two geometries, the predictions converge to a similarly precise relation (at least when $L$ is order one):
 \begin{itemize} 
 \item For the internal symmetry based on the hyperbolic geometry, one is led to
  \begin{align} \label{alphaplot}
    n_s = 1 - \frac2N + \mathcal{O}\left(\frac{1}{N^2}\right) \,, \quad r = \frac{2 L^2}{N^2} + \mathcal{O}\left(\frac{1}{N^3}\right) \,.
\end{align}
The spectral index has a model-independent leading term, while subleading terms are sufficiently suppressed. Moreover, the $1/N^2$ scaling leads to permille values of the tensors, indicated by the dashed lines. Note that these only depend on $L$. As a consequence, the $L=1$ points (denoted by blue dots) for all monomials coincide.
 \item In the case of the spacetime symmetry based on the AdS geometry, one finds 
  \begin{align} \label{adsplot}
 n_s = 1 - \frac{4}{3N} - \frac{(mL)^{2/3}}{(3N)^{4/3}} + \mathcal{O}\left(\frac{1}{N^{5/3}}\right) \,, \quad r = \frac{8 (mL)^{2/3}}{(3N)^{4/3}} + \mathcal{O}\left(\frac{1}{N^2}\right).
  \end{align}
While the leading term for the spectral index is again model-independent, there is a subleading term which is only $1/N^{1/3}$ suppressed and hence can make a (possibly observable) model-dependent contribution. Similarly, the $1/ N^{4/3}$ leads to percent values for the tensors indicated by the solid lines. Their value in this case depends on $mL$. Therefore the $L=1$ points differ slightly in their $(n_s,r)$ values.
\end{itemize}
These can be seen as prototypes for chaotic inflation, $\alpha$-attractors and ambient inflation, respectively. The robustness of the latter two is beautifully illustrated by the funnel-like behaviour in the logarithmic $(n_s,r)$ plane.

For intermediate values of $L$ one finds an interpolation between the different cases. Note that these do not conform to either of the $1/N$ scalings introduced in the previous subsection. The reason for this is the interplay between the large values of $N$ and $L$ in these cases: while certain terms might be higher-order in $1/N$, their $L$-prefactor might offset this and make them equally important. We are not aware of any simple expansion or organising principle in this intermediate regime. Moreover, in this regime the model-independence is lost; for e.g.~$\alpha$-attractors there are different interpolations between the regimes of $L$ infinite and of order one.

Note that there are specific flat limits which have the same spectral index as either of the curved models as we decrease $L$. The corresponding models can be extracted by equating the spectral indices in \eqref{mono-predictions} and \eqref{pole-predictions}, leading to the relation $m = 2/(p-1)$ between the pole $p$ of the curved model and the power $m$ of the corresponding monomial. For $\alpha$-attractors this is the quadratic potential. In a sense, this model can therefore be seen as the simplest flat counterpart to the $\alpha$-attractors: as one varies $L$, the spectral index is invariant while the tensor-to-scalar ratio interpolates from a $1/N^2$ to a $1/N$ scaling. In contrast, for ambient inflation the special flat limit yields a $\phi^{2/3}$ behaviour. Again, as one varies $L$, the spectral index is roughly invariant while the $1/N$ scaling of $r$ varies from $4/3$ to $1$. It is tantalising to notice that exactly this fractional behaviour surfaced in the first realisation of axion monodromy inflation \cite{monodromy1}. One might hope that there is an extension of this scenario where one includes the curvature scale $L$ and obtains a microscopic realisation of the ambient inflation scenario.

\section{Conclusions}
The recent CMB data indicates that the most simple single field inflationary models do not seem to describe the physics of the very early universe, since they predict a tensor-to-scalar ratio above the current upper bound \cite{planck,bicep}. This requires model builders to construct slightly more complicated inflationary models but one should aim to maintain the very nice field theory properties of these original models, namely, radiative stability within EFT. This is crucial to ensure that slow-roll inflation can take place while perturbative quantum corrections are under control. For simple single field models this is the case thanks to a weakly broken shift symmetry in the kinetic sector.

Multi-field inflation is a popular alternative to single field models given that high-energy frameworks such as string theory and supergravity often involve a large number of moduli. In this paper we have constructed kinetic sectors for multi-field inflation which have a non-linearly realised symmetry which can be weakly broken by a potential to drive inflation. This in direct comparison to what happens in simple chaotic inflation. Indeed, throughout our motivation has been to build observationally consistent inflationary models which maintain radiative stability within EFT.  

We have classified possible kinetic structures for $n$ scalar fields which are fixed by a non-linearly realised symmetry corresponding to a coset space $G/H$ similar to what was done in \cite{cliffcosets}. Concentrating on cases where $G$ is maximally symmetric, five different kinetic sectors are possible and each come with their own interesting structures. Three of these arise when $G$ is an internal group and the other two arise when it is a space-time extension of the 4-dimensional Poincar\'{e} group. In all cases we made use of the coset construction to extract the invariant kinetic sectors and for the space-time symmetric cases we imposed various inverse Higgs constraints to arrive at a theory of interacting scalar fields. Out of these five possibilites our main focus has been on two examples which when one adds the symmetry breaking potential correspond to the well known $\alpha$-attractors model and a new inflationary model we called ambient inflation. In the two field limit these models describe different interactions between an axion and a dilaton. 

Although $\alpha$-attractors and ambient inflation have a comparable origin with their kinetic sectors dictated by symmetry, their inflationary predictions differ. The spectral index for $\alpha$-attractors naturally lies close to the point $n_{s} = 0.965$ whereas for ambient inflation one finds predictions closer to $n_{s} = 0.975$. A similar difference can be seen in the tensor-to-scalar ratios which naturally differ by an order of magnitude due their $1/N^2$ and $1/N^{4/3}$ scaling for order one parameters. These lead to a range of permille and percent level values for the tensors, well within the reach of future CMB missions. 

While the present discussion has highlighted the symmetries of the bosonic scalar sectors, it is natural to embed these in more formal constructions in order to connect to high-energy theories. In this context, $\alpha$-attractors arise very naturally in scenarios with minimal supersymmetry or superconformal symmetry \cite{Porrati, alpha-attractors} as well as no-scale supergravity \cite{Ellis1, Ellis2}. Such set-ups often have a K\"{a}hler potential with specific symmetries which are weakly broken by the superpotential (alternatively, the symmetry breaking can also be realized in terms of the K\"{a}hler potential, indicating an interesting link with anti-D3 brane geometry \cite{Scalisi, Yamada}). Moreover, there are proposals to realise the same inflationary scenarios in maximally supersymmetric theories as well as string theory \cite{Ferrara:2016, Wrase, Burgess:2016, Fibre}. For the new  bosonic construction, based on the Anti-de Sitter ambient space, it would be very interesting to investigate similar set-ups. In particular, this would involve the spontaneous breaking of the 6-dimensional Anti-de Sitter superalgebra with eight supercharges to 4-dimensional minimal super-Poincar\'{e}, which can be seen as the curved version of DBI-Volkov-Akulov \cite{Rocek}.

On a final note, while in this paper our non-linear symmetries were weakly broken by the inflationary potential, it would also be interesting to consider the effects of unbroken non-linearly realised symmetries on cosmological correlators both in the single field and multi-field case. This is the cosmological version of studying the effects of non-linear symmetries in the soft limits of scattering amplitudes \cite{specialgal1,periodic,PSW} as we discussed in the introduction. For a single field with a shift symmetry this was very recently explored in \cite{shiftcosmo1, shiftcosmo2} and novel features can arise. Based on what we know about scattering amplitudes, one would expect new novel features to arise if one considers space-time extensions of this shift symmetry. This is an interesting avenue for future work.

\section*{Acknowledgements}

We would like to thank Marco Boers, Garrett Goon, Renata Kallosh, Nemanja Kaloper, Andrei Linde, Antonio Padilla, Paul Townsend and Pelle Werkman for useful discussions. The authors acknowledge the Dutch funding agency ‘Netherlands Organisation for Scientific Research’ (NWO) for financial support.


\begin{thebibliography}{99}

\bibitem{adler}
  S.~L.~Adler,
  Phys.\ Rev.\  {\bf 137} (1965) B1022.

\bibitem{specialgal1}
  C.~Cheung, K.~Kampf, J.~Novotny and J.~Trnka,
  Phys.\ Rev.\ Lett.\  {\bf 114} (2015) no.22,  221602
  [arXiv:1412.4095 [hep-th]].

\bibitem{specialgal2}
  K.~Hinterbichler and A.~Joyce,
  Phys.\ Rev.\ D {\bf 92} (2015) no.2,  023503
  [arXiv:1501.07600 [hep-th]].

\bibitem{periodic}
  C.~Cheung, K.~Kampf, J.~Novotny, C.~H.~Shen and J.~Trnka,
  JHEP {\bf 1702} (2017) 020
  [arXiv:1611.03137 [hep-th]].

\bibitem{PSW}
  A.~Padilla, D.~Stefanyszyn and T.~Wilson,
  JHEP {\bf 1704} (2017) 015
  [arXiv:1612.04283 [hep-th]].

\bibitem{chaotic}
  A.~D.~Linde,
  Phys.\ Lett.\  {\bf 129B} (1983) 177.

\bibitem{quantumcorrections}
  A.~D.~Linde,
  Phys.\ Lett.\ B {\bf 202} (1988) 194.

\bibitem{KLS}
  N.~Kaloper, A.~Lawrence and L.~Sorbo,
  JCAP {\bf 1103} (2011) 023
  [arXiv:1101.0026 [hep-th]].

\bibitem{planck}
  P.~A.~R.~Ade {\it et al.} [Planck Collaboration],
  Astron.\ Astrophys.\  {\bf 594} (2016) A20
  [arXiv:1502.02114 [astro-ph.CO]].

\bibitem{bicep}
  P.~A.~R.~Ade {\it et al.} [BICEP2 and Keck Array Collaborations],
  Phys.\ Rev.\ Lett.\  {\bf 116} (2016) 031302
  [arXiv:1510.09217 [astro-ph.CO]].

\bibitem{cliffcosets}
  C.~P.~Burgess, M.~Cicoli, F.~Quevedo and M.~Williams,
  JCAP {\bf 1411} (2014) 045
  [arXiv:1404.6236 [hep-th]].

\bibitem{zoo}
  A.~Nicolis, R.~Penco, F.~Piazza and R.~Rattazzi,
  JHEP {\bf 1506} (2015) 155
  [arXiv:1501.03845 [hep-th]].

\bibitem{paul}
  J.~Sonner and P.~K.~Townsend,
  Class.\ Quant.\ Grav.\  {\bf 23} (2006) 441
  [hep-th/0510115].


\bibitem{internal1}
  S.~R.~Coleman, J.~Wess and B.~Zumino,
  Phys.\ Rev.\  {\bf 177} (1969) 2239.

\bibitem{internal2}
  C.~G.~Callan, Jr., S.~R.~Coleman, J.~Wess and B.~Zumino,
  Phys.\ Rev.\  {\bf 177} (1969) 2247.

\bibitem{spacetime1}
  D.~V.~Volkov,
  Fiz.\ Elem.\ Chast.\ Atom.\ Yadra {\bf 4} (1973) 3.

\bibitem{spacetime2}
  E.~A.~Ivanov and V.~I.~Ogievetsky,
  Teor.\ Mat.\ Fiz.\  {\bf 25} (1975) 164.

\bibitem{EFTofI1}
  C.~Cheung, P.~Creminelli, A.~L.~Fitzpatrick, J.~Kaplan and L.~Senatore,
  JHEP {\bf 0803} (2008) 014
  [arXiv:0709.0293 [hep-th]].

\bibitem{EFTofI2}
  L.~Senatore and M.~Zaldarriaga,
  JHEP {\bf 1204} (2012) 024
  [arXiv:1009.2093 [hep-th]].

\bibitem{alpha-attractors}
  R.~Kallosh, A.~Linde and D.~Roest,
  JHEP {\bf 1311} (2013) 198
  [arXiv:1311.0472 [hep-th]].

\bibitem{alpha-attractors-corrections}
  R.~Kallosh and A.~Linde,
  JCAP {\bf 1606} (2016) no.06,  047
  [arXiv:1604.00444 [hep-th]].

\bibitem{globalbreaking}
  R.~Kallosh, A.~D.~Linde, D.~A.~Linde and L.~Susskind,
  Phys.\ Rev.\ D {\bf 52} (1995) 912
  [hep-th/9502069].

\bibitem{monodromy1}
  E.~Silverstein and A.~Westphal,
  Phys.\ Rev.\ D {\bf 78} (2008) 106003
  [arXiv:0803.3085 [hep-th]].

\bibitem{monodromy2}
  L.~McAllister, E.~Silverstein and A.~Westphal,
  Phys.\ Rev.\ D {\bf 82} (2010) 046003
  [arXiv:0808.0706 [hep-th]].

\bibitem{KS}
  N.~Kaloper and L.~Sorbo,
  Phys.\ Rev.\ Lett.\  {\bf 102} (2009) 121301
  [arXiv:0811.1989 [hep-th]].



\bibitem{inferno}
  M.~Berg, E.~Pajer and S.~Sjors,
  Phys.\ Rev.\ D {\bf 81} (2010) 103535
  [arXiv:0912.1341 [hep-th]].

\bibitem{flatten}
  X.~Dong, B.~Horn, E.~Silverstein and A.~Westphal,
  Phys.\ Rev.\ D {\bf 84} (2011) 026011
  [arXiv:1011.4521 [hep-th]].

\bibitem{london}
  N.~Kaloper and A.~Lawrence,
  Phys.\ Rev.\ D {\bf 95} (2017) no.6,  063526
  [arXiv:1607.06105 [hep-th]].

\bibitem{fourpi}
  G.~D'Amico, N.~Kaloper and A.~Lawrence,
  arXiv:1709.07014 [hep-th].

\bibitem{KRS}
  R.~Klein, D.~Roest and D.~Stefanyszyn,
  JHEP {\bf 1710} (2017) 051
  [arXiv:1709.03525 [hep-th]].


\bibitem{wess}
  G.~Goon, K.~Hinterbichler, A.~Joyce and M.~Trodden,
  JHEP {\bf 1206} (2012) 004
  [arXiv:1203.3191 [hep-th]].

\bibitem{higher}
  K.~Hinterbichler, M.~Trodden and D.~Wesley,
  Phys.\ Rev.\ D {\bf 82} (2010) 124018
  [arXiv:1008.1305 [hep-th]].

\bibitem{reunited}
  C.~de Rham and A.~J.~Tolley,
  JCAP {\bf 1005} (2010) 015
  [arXiv:1003.5917 [hep-th]].

\bibitem{equivalence}
  S.~Bellucci, E.~Ivanov and S.~Krivonos,
  Phys.\ Rev.\ D {\bf 66} (2002) 086001
   Erratum: [Phys.\ Rev.\ D {\bf 67} (2003) 049901]
  [hep-th/0206126].


\bibitem{mapping}
  P.~Creminelli, M.~Serone and E.~Trincherini,
  JHEP {\bf 1310} (2013) 040
  [arXiv:1306.2946 [hep-th]].


\bibitem{dbiinflation}
  M.~Alishahiha, E.~Silverstein and D.~Tong,
  Phys.\ Rev.\ D {\bf 70} (2004) 123505
  [hep-th/0404084].

\bibitem{nemanjascale}
  C.~Csaki, N.~Kaloper, J.~Serra and J.~Terning,
  Phys.\ Rev.\ Lett.\  {\bf 113} (2014) 161302
  [arXiv:1406.5192 [hep-th]].

\bibitem{alphageometry}
  J.~J.~M.~Carrasco, R.~Kallosh, A.~Linde and D.~Roest,
  Phys.\ Rev.\ D {\bf 92} (2015) no.4,  041301
  [arXiv:1504.05557 [hep-th]].

\bibitem{rubio}
  G.~K.~Karananas and J.~Rubio,
  Phys.\ Lett.\ B {\bf 761} (2016) 223
  doi:10.1016/j.physletb.2016.08.037
  [arXiv:1606.08848 [hep-ph]].

\bibitem{Garcia} 
  J.~Garcia-Bellido and D.~Roest,
  Phys.\ Rev.\ D {\bf 89}, no. 10, 103527 (2014)
  [arXiv:1402.2059 [astro-ph.CO]].

\bibitem{Freese} 
  K.~Freese, J.~A.~Frieman and A.~V.~Olinto,
  Phys.\ Rev.\ Lett.\  {\bf 65}, 3233 (1990).

\bibitem{universality}
  D.~Roest,
  JCAP {\bf 1401} (2014) 007
  [arXiv:1309.1285 [hep-th]].

\bibitem{unity}
  M.~Galante, R.~Kallosh, A.~Linde and D.~Roest,
  Phys.\ Rev.\ Lett.\  {\bf 114} (2015) no.14,  141302
  [arXiv:1412.3797 [hep-th]].

\bibitem{pole}
  B.~J.~Broy, M.~Galante, D.~Roest and A.~Westphal,
  JHEP {\bf 1512} (2015) 149
  [arXiv:1507.02277 [hep-th]].

\bibitem{contour}
  T.~Terada,
  Phys.\ Lett.\ B {\bf 760} (2016) 674
  [arXiv:1602.07867 [hep-th]].

\bibitem{Porrati} 
  S.~Ferrara, R.~Kallosh, A.~Linde and M.~Porrati,
  Phys.\ Rev.\ D {\bf 88}, no. 8, 085038 (2013)
  [arXiv:1307.7696 [hep-th]].

\bibitem{Ferrara:2016} 
  S.~Ferrara and R.~Kallosh,
  Phys.\ Rev.\ D {\bf 94}, no. 12, 126015 (2016)
  [arXiv:1610.04163 [hep-th]].

\bibitem{Wrase} 
  R.~Kallosh, A.~Linde, T.~Wrase and Y.~Yamada,
  JHEP {\bf 1704}, 144 (2017)
  [arXiv:1704.04829 [hep-th]].


\bibitem{WGC1}
  N.~Arkani-Hamed, L.~Motl, A.~Nicolis and C.~Vafa,
  JHEP {\bf 0706} (2007) 060
  [hep-th/0601001].

\bibitem{WGC2}
  J.~Brown, W.~Cottrell, G.~Shiu and P.~Soler,
  JHEP {\bf 1510} (2015) 023
  [arXiv:1503.04783 [hep-th]].

\bibitem{WGC3}
  D.~Klaewer and E.~Palti,
  JHEP {\bf 1701} (2017) 088
  [arXiv:1610.00010 [hep-th]].

\bibitem{doublealpha}
  A.~Achúcarro, R.~Kallosh, A.~Linde, D.~G.~Wang and Y.~Welling,
  arXiv:1711.09478 [hep-th].

\bibitem{Ellis1}
  J.~Ellis, D.~V.~Nanopoulos and K.~A.~Olive,
  Phys.\ Rev.\ Lett.\  {\bf 111} (2013) 111301
   Erratum: [Phys.\ Rev.\ Lett.\  {\bf 111} (2013) no.12,  129902]
  [arXiv:1305.1247 [hep-th]].

\bibitem{Ellis2} 
  J.~Ellis, D.~V.~Nanopoulos and K.~A.~Olive,
  JCAP {\bf 1310}, 009 (2013)
  [arXiv:1307.3537 [hep-th]].

\bibitem{Scalisi}
  E.~McDonough and M.~Scalisi,
  JCAP {\bf 1611}, no. 11, 028 (2016)
  [arXiv:1609.00364 [hep-th]].

\bibitem{Yamada} 
  R.~Kallosh, A.~Linde, D.~Roest and Y.~Yamada,
  JHEP {\bf 1707}, 057 (2017)
  [arXiv:1705.09247 [hep-th]].



\bibitem{Burgess:2016} 
  C.~P.~Burgess, M.~Cicoli, S.~de Alwis and F.~Quevedo,
  JCAP {\bf 1605}, no. 05, 032 (2016)
  [arXiv:1603.06789 [hep-th]].

\bibitem{Fibre} 
  R.~Kallosh, A.~Linde, D.~Roest, A.~Westphal and Y.~Yamada,
  arXiv:1707.05830 [hep-th].


\bibitem{Rocek} 
  M.~Rocek and A.~A.~Tseytlin,
  Phys.\ Rev.\ D {\bf 59}, 106001 (1999)
  [hep-th/9811232].

\bibitem{shiftcosmo1}
  R.~Bravo, S.~Mooij, G.~A.~Palma and B.~Pradenas,
  arXiv:1711.02680 [astro-ph.CO].


\bibitem{shiftcosmo2}
  B.~Finelli, G.~Goon, E.~Pajer and L.~Santoni,
  arXiv:1711.03737 [hep-th].








\end{thebibliography}
\end{document}